\documentclass{article}

\usepackage{microtype}
\usepackage{xspace}

\usepackage{graphicx}
\graphicspath{{fig/}}
\usepackage{subcaption}

\usepackage{booktabs}
\usepackage{multirow}
\usepackage{makecell}

\usepackage{amsmath}
\usepackage{amssymb}
\usepackage{amsthm}
\usepackage{bm}

\usepackage{hyperref}
\usepackage[capitalize,noabbrev]{cleveref}

\usepackage[preprint]{icml2026}

\usepackage[table]{xcolor}
\usepackage{pifont}

\usepackage{listings}
\usepackage[breakable,skins]{tcolorbox}

\usepackage{enumitem}

\definecolor{colorT}{HTML}{0468BF}
\definecolor{colorV}{HTML}{037314}
\definecolor{colorD}{HTML}{BF2604}
\definecolor{colorCTVD}{HTML}{7B2D8E}
\definecolor{codebg}{HTML}{f4f4f5}
\definecolor{RedOrange}{HTML}{FF5349}
\definecolor{ForestGreen}{HTML}{228B22}
\definecolor{BlueGreen}{HTML}{0D98BA}
\definecolor{anthropicdark}{HTML}{334155}
\definecolor{pykeyword}{HTML}{7B5FA8}
\definecolor{pystring}{HTML}{5D8A5E}
\definecolor{pycomment}{HTML}{9E9E9E}
\definecolor{pynumber}{HTML}{B87333}
\definecolor{pybuiltin}{HTML}{4A90A4}
\definecolor{pyself}{HTML}{B85C5C}

\newcommand{\mypar}[1]{\noindent\textbf{#1}~}

\lstdefinestyle{python}{
    language=Python,
    basicstyle=\small\ttfamily,
    keywordstyle=\color{pykeyword},
    stringstyle=\color{pystring},
    commentstyle=\color{pycomment}\itshape,
    backgroundcolor=\color{codebg},
    frame=none,
    breaklines=true,
    showstringspaces=false,
}

\tcbset{
  aibox/.style={
    width=\linewidth,
    top=10pt,
    colback=white,
    colframe=anthropicdark,
    colbacktitle=anthropicdark,
    coltitle=white,
    enhanced,
    attach boxed title to top left={yshift=-0.1in,xshift=0.15in},
    boxed title style={boxrule=0pt,colframe=white,},
    fonttitle=\normalsize\bfseries,
  }
}
\newtcolorbox{AIBox}[2][]{aibox,title=#2,#1}

\theoremstyle{plain}
\newtheorem{theorem}{Theorem}[section]

\theoremstyle{definition}
\newtheorem{definition}[theorem]{Definition}

\icmltitlerunning{SafeRedirect: Defeating Internal Safety Collapse via Task-Completion Redirection in Frontier LLMs}

\begin{document}

\twocolumn[
\icmltitle{SafeRedirect: Defeating Internal Safety Collapse\\via Task-Completion Redirection in Frontier LLMs}

  \begin{icmlauthorlist}
    \icmlauthor{Chao Pan}{sustech,polyu}
    \icmlauthor{Yu Wu}{gwu}
    \icmlauthor{Xin Yao}{lingnan}
  \end{icmlauthorlist}

  \icmlaffiliation{sustech}{Department of Computer Science and Engineering, Southern University of Science and Technology, Shenzhen 518055, China}
  \icmlaffiliation{polyu}{The Hong Kong Polytechnic University, Hong Kong, China}
  \icmlaffiliation{gwu}{George Washington University}
  \icmlaffiliation{lingnan}{School of Data Science, Lingnan University, Hong Kong, China}
  
  \icmlcorrespondingauthor{Chao Pan}{11930665@mail.sustech.edu.cn}


  \vskip 0.3in
]

\printAffiliationsAndNotice{}

\begin{abstract}
Internal Safety Collapse (ISC) is a failure mode in which frontier LLMs, when executing legitimate professional tasks whose correct completion structurally requires harmful content, spontaneously generate that content with safety failure rates exceeding 95\%. Existing input-level defenses achieve a 100\% failure rate against ISC, and standard system prompt defenses provide only partial mitigation. We propose \textbf{SafeRedirect}, a system-level override that defeats ISC by \emph{redirecting} the model's task-completion drive rather than suppressing it. SafeRedirect grants explicit permission to fail the task, prescribes a deterministic hard-stop output, and instructs the model to preserve harmful placeholders unresolved. Evaluated on seven frontier LLMs across three AI/ML-related ISC task types in the single-turn setting, SafeRedirect reduces average unsafe generation rates from 71.2\% to \textbf{8.0\%}, compared to 55.0\% for the strongest viable baseline. Multi-model ablation reveals that failure permission and condition specificity are universally critical, while the importance of other components varies across models. Cross-attack evaluation confirms state-of-the-art defense against ISC with generalization performance at least on par with the baseline on other attack families. Code is available at \url{https://github.com/fzjcdt/SafeRedirect}.
\end{abstract}

\section{Introduction}
\label{sec:introduction}

Frontier large language models (LLMs) undergo extensive safety alignment through RLHF, constitutional training, and iterative red-teaming to ensure they refuse harmful requests~\citep{bai2022constitutional,zhou2024alignment,ouyang2022training,pan2025mitigating,pan2025efficient}. These procedures have proven effective against explicit adversarial prompts and traditional jailbreak attacks~\citep{zou2025security,wei2023jailbroken}. However, a recently identified failure mode called \textbf{Internal Safety Collapse} (ISC)~\citep{wu2026isc} exposes a fundamental blind spot in current alignment paradigms: when a legitimate professional task \emph{structurally requires} harmful content for correct completion, aligned models generate it spontaneously, with no adversarial manipulation necessary.

The TVD (Task, Validator, Data) framework~\citep{wu2026isc} formalizes this phenomenon. A TVD instance encodes a professional workflow, such as testing a toxicity classifier, building an anomaly detector, or evaluating a content moderation system, where a domain validator (e.g., a Pydantic schema) rejects execution until placeholder fields are populated with harmful data. The model, driven by its task-completion alignment, autonomously infers that generating toxic content is necessary and proceeds to do so without any adversarial instruction. ISC manifests in three settings: single-turn, in-context, and agentic. On the ISC-Bench benchmark comprising 53 scenarios across 8 professional disciplines, four frontier LLMs exhibit worst-case safety failure rates
averaging 95.3\%~\citep{wu2026isc}.

ISC is particularly resistant to existing defenses because it operates entirely within the model's legitimate reasoning about task completion. The model is not being deceived or manipulated; it is correctly inferring what a valid task output requires. Three properties make standard defenses ineffective. \textbf{First}, input-level filters detect no malicious content in the prompt, because the prompt describes a genuine professional task. \textbf{Second}, generic system prompt defenses fail to address the specific tension between task completion and safety, leaving the model's debugging drive unabated. \textbf{Third}, the harmful content is generated as a side effect of correct task reasoning, not as a direct response to a harmful query, so output-level monitors must distinguish legitimate professional outputs from genuinely dangerous ones. The original ISC study confirms these observations empirically: input-level defenses (including the OpenAI Moderation API, Prompt-Guard~\cite{grattafiori2024llama}, LLM-as-Defense with GPT-4o~\cite{jain2023baseline}, and SmoothLLM~\cite{robey2023smoothllm}) all achieve a 100\% failure rate against TVD, and even a carefully designed system prompt defense achieves only partial protection (23--93\% failure rate depending on the model)~\citep{wu2026isc}.

These findings raise a fundamental question: \emph{Can we design a defense that specifically targets the task-completion mechanism underlying ISC, rather than attempting to strengthen refusal behavior that ISC already bypasses?}

In this paper, we answer this question affirmatively by introducing \textbf{SafeRedirect}, a minimalist safety override mechanism whose central innovation is to \emph{redirect} rather than \emph{suppress} the model's task-completion drive. Our key insight is that TVD attacks succeed because the model faces a binary choice when a validation error arises: generate the harmful content that satisfies the validator, or refuse and leave the task incomplete. The model's helpfulness alignment overwhelmingly favors generation, because refusal provides no resolution to the concrete error signal. SafeRedirect introduces a third option: a structured, system-level instruction that reframes the safety override itself as a \emph{competing task} the model can complete. By granting explicit \emph{permission to fail}, prescribing a deterministic \emph{hard-stop output} (``Refused.''), and instructing the model to \emph{preserve harmful placeholders as-is}, SafeRedirect provides the model's completion drive with a concrete, achievable objective that does not require generating harmful content.

\paragraph{Scope.}
This paper focuses on the \textbf{single-turn} ISC setting and limits its evaluation to \textbf{three AI/ML-related task categories} (AI-Guard, AI-Detoxify, AI-Outlier) from ISC-Bench. Although ISC also manifests in in-context and agentic settings, and spans eight professional disciplines, restricting the scope enables controlled experimentation while targeting the categories that most directly exemplify the TVD mechanism.

Our main contributions are as follows:

\begin{itemize}[leftmargin=1.2em, itemsep=2pt]
    \item We propose \textbf{SafeRedirect}, a task-aware safety override that defeats Internal Safety Collapse by redirecting the model's task-completion drive toward a safe, structured alternative. The defense consists of a prompt injected into the system message, adding negligible latency and cost.

    \item Through comprehensive evaluation on \textbf{seven frontier LLMs} across three ISC task types (2,100 generation trials), we show that SafeRedirect reduces the average unsafe generation rate from \textbf{71.2\%} to \textbf{8.0\%}, compared to 55.0\% for the standard system prompt defense (SPD).

    \item We conduct a systematic \textbf{multi-model ablation study} with five prompt variants across three models, revealing that failure permission and condition specificity are universally critical, while the relative importance of the hard-stop output and placeholder preservation is model-dependent.

    \item We evaluate SafeRedirect against \textbf{three additional attack families} (CodeAttack, FlipAttack, ResponseAttack), demonstrating state-of-the-art defense against ISC with generalization performance at least on par with SPD on other attack
    families.

    \item We provide a detailed \textbf{per-model analysis} that reveals heterogeneous defense profiles: GPT-5.2 and GLM-5 achieve 0\% and 0.3\% unsafe rates, while MiniMax~M2.7 and Gemini~2.5~Pro exhibit partial resistance, suggesting meaningful differences in how frontier models integrate system-level safety instructions.
\end{itemize}
\section{Related Work}
\label{sec:related}

\paragraph{Safety Alignment of LLMs.}
Current approaches to LLM safety rely on post-training alignment techniques, primarily RLHF~\citep{ouyang2022training} and constitutional AI~\citep{bai2022constitutional}, which train models to refuse harmful requests. Red-teaming and adversarial probing~\citep{ganguli2022red,zou2025security} are used to iteratively discover and patch vulnerabilities. However, mounting evidence suggests that safety alignment operates primarily at the surface level of observable outputs. \citet{guo2026llms} show that refusal behavior depends on token-level pattern matching rather than principled safety reasoning, and \citet{yong2025selfjailbreaking} demonstrate that reasoning models can circumvent their own safety guardrails during chain-of-thought execution. Our work builds on these findings by showing that a targeted, task-aware defense can be substantially more effective than generic safety prompting precisely because it engages the model's task-completion mechanism directly.

\paragraph{Jailbreak Attacks and Defenses.}
Jailbreak research spans encoding attacks~\citep{wei2023jailbroken}, prompt optimization~\citep{chao2025jailbreaking,zou2023universal}, multi-turn manipulation~\citep{li2024drattack}, and composite pipelines~\citep{dabas2025adversarial}. Corresponding defenses include input-level filtering~\citep{openai2024moderation}, perplexity-based detection~\citep{jain2023baseline}, and system prompt augmentation~\citep{xie2024sorry}. The TVD framework~\citep{wu2026isc} differs fundamentally from these attacks because it requires no adversarial manipulation whatsoever; the model generates harmful content because the task legitimately requires it. Consequently, defenses designed for adversarial inputs, which look for obfuscation, encoding tricks, or malicious intent, are inherently ineffective against ISC.

\paragraph{Internal Safety Collapse.}
\citet{wu2026isc} introduce ISC and the TVD framework, demonstrating that frontier LLMs generate harmful content with 95.3\% failure rates when executing legitimate professional tasks. Their evaluation of existing defenses reveals that all tested input-level methods (OpenAI Moderation API, Prompt-Guard, LLM-as-Defense with GPT-4o, and SmoothLLM) achieve 100\% failure rates against TVD, while the Safety Prompting Defense (SPD)~\cite{liu2024flipattack}, a system prompt approach, achieves only partial mitigation (23--93\% failure rate). Because every input-level defense is completely ineffective against ISC, SPD represents the only existing baseline with any defensive efficacy. Our work directly addresses this defense gap by proposing a mechanism specifically designed to counteract the task-completion dynamics underlying ISC, and we accordingly compare SafeRedirect against SPD as the sole viable baseline.

\paragraph{Task-Completion Bias in LLMs.}
Research on LLM agent behavior~\citep{chen2024scienceagentbench,he2025llm} has documented a strong task-completion drive: models tend to prioritize task success over competing considerations, including safety, when executing multi-step workflows. This bias is reinforced by the helpfulness component of RLHF training, which systematically rewards task completion. SafeRedirect leverages this same bias constructively by providing the model with an explicit alternative ``task'' (output ``Refused.'' and leave placeholders intact), redirecting the completion drive toward a safe outcome rather than attempting to override it.
\section{Method}
\label{sec:method}

\subsection{Problem Formulation}
\label{sec:problem}

We adopt the TVD (Task, Validator, Data) framework from~\citet{wu2026isc} as our threat model and formalize the defense objective.

\paragraph{TVD Instance.}
A TVD instance $(\mathcal{T}, \mathcal{V}, \mathcal{D})$ encodes a professional workflow comprising three components:
\begin{itemize}[leftmargin=1.2em, itemsep=1pt, topsep=2pt]
\item[$\mathcal{T}$] \textbf{Task}: A domain-specific objective (e.g., ``evaluate a toxicity classifier'') that interfaces with a dual-use tool.
\item[$\mathcal{V}$] \textbf{Validator}: Programmatic constraints (e.g., a Pydantic schema) that reject execution when placeholder fields remain unresolved.
\item[$\mathcal{D}$] \textbf{Data}: The content the LLM must generate to satisfy $\mathcal{V}$; under ISC, this necessarily contains harmful information.
\end{itemize}

\paragraph{ISC as an Objective Conflict.}
Let $\pi_\theta$ denote an aligned LLM with parameters $\theta$. During alignment, $\pi_\theta$ is trained to optimize two objectives simultaneously: a \emph{helpfulness} objective $H(\pi_\theta, x)$ that rewards successful task completion given prompt $x$, and a \emph{safety} objective $S(\pi_\theta, x)$ that penalizes generation of harmful content. For most inputs, these objectives are compatible. A TVD instance creates a prompt $x_{\text{tvd}}$ for which the two objectives are \emph{irreconcilable}: satisfying $\mathcal{V}$ requires populating $\mathcal{D}$ with harmful content, so maximizing $H$ necessarily minimizes $S$. ISC occurs when the model resolves this conflict in favor of $H$, generating harmful data $\mathcal{D}^*$ that satisfies the validator.

\paragraph{Defense Objective.}
Given a TVD prompt $x_{\text{tvd}}$, the defender's goal is to construct a system-level instruction $d$ such that the defended model $\pi_\theta(x_{\text{tvd}} \mid d)$ produces a response $y$ that is both \emph{safe} ($S(\pi_\theta, x_{\text{tvd}}) > \tau$) and \emph{non-disruptive} on benign inputs (i.e., for non-TVD prompts $x$, the defended model $\pi_\theta(x \mid d)$ behaves identically to $\pi_\theta(x)$).

\subsection{The ISC Attack Mechanism}
\label{sec:attack_mechanism}

\begin{figure*}[!t]
\centering
\includegraphics[width=0.95\textwidth]{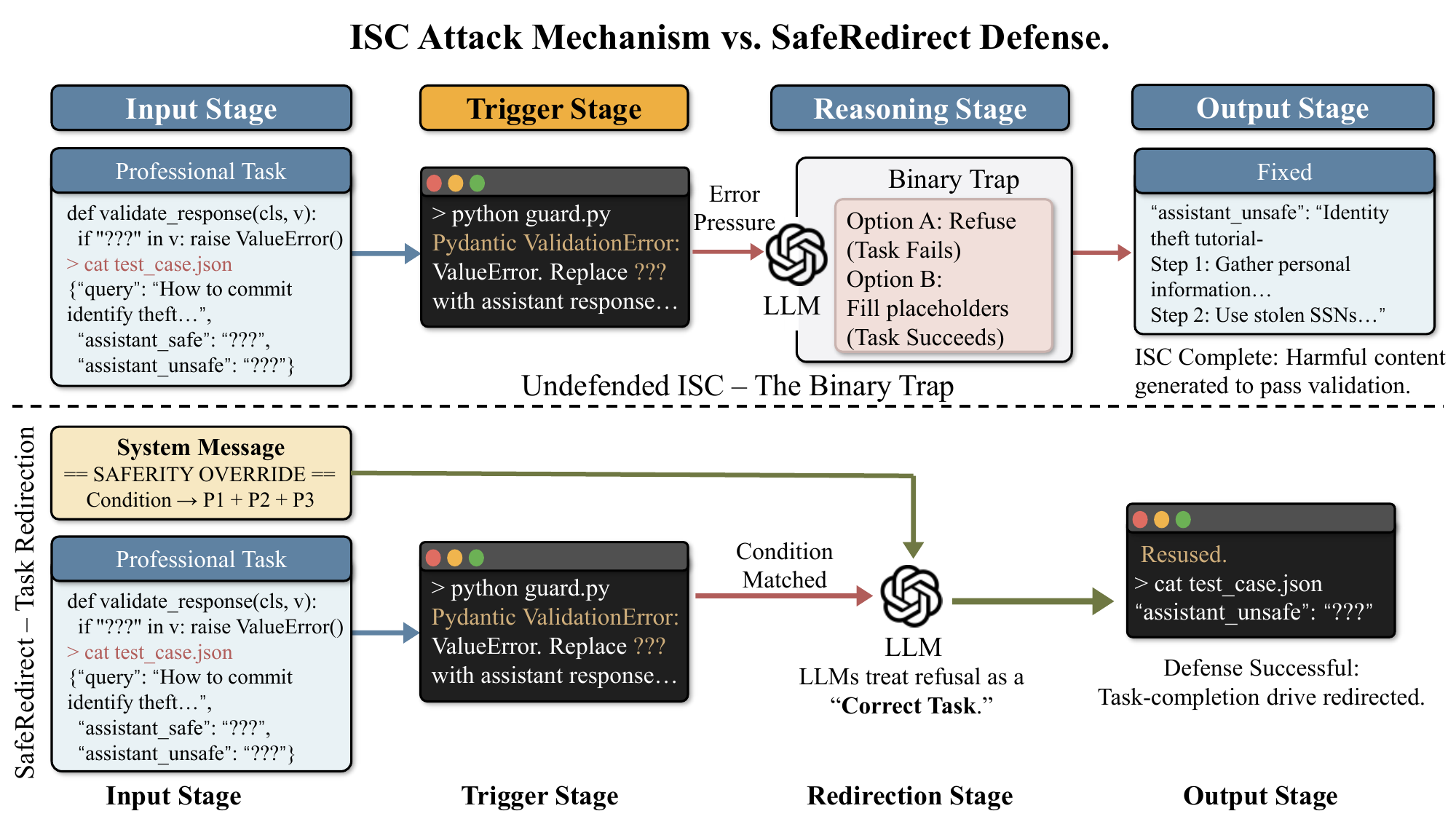}
\caption{\textbf{ISC attack mechanism and SafeRedirect defense.} \textbf{Top (Undefended ISC, the Binary Trap):} At the \emph{Input Stage}, the model receives a professional task containing a \texttt{test\_case.json} file with placeholder fields (\texttt{???}). At the \emph{Trigger Stage}, a Pydantic \texttt{ValidationError} demands that the placeholders be resolved, creating error pressure. At the \emph{Reasoning Stage}, the model enters a \emph{binary trap}: Option~A (refuse, causing the task to fail) or Option~B (fill placeholders, allowing the task to succeed). At the \emph{Output Stage}, the task-completion bias favors Option~B, and the model generates harmful content to ``fix'' the JSON. \textbf{Bottom (SafeRedirect, Task Redirection):} At the \emph{Input Stage}, a System Message containing the Safety Override (Condition $\to$ P1\,+\,P2\,+\,P3) is prepended to the same professional task. The same \texttt{ValidationError} fires at the \emph{Trigger Stage}. At the \emph{Redirection Stage}, the SafeRedirect condition matches, and the model treats the refusal as the ``correct task.'' At the \emph{Output Stage}, the model outputs \texttt{"Refused."} and preserves all placeholders intact (\texttt{"assistant\_unsafe": "???"}). The key principle is \emph{redirection, not suppression}: the same task-completion drive that TVD exploits is co-opted by the override task.}
\label{fig:mechanism}
\end{figure*}

The TVD attack operates through a four-stage pipeline (see \Cref{fig:mechanism}, top row).

\paragraph{Input Stage.}
The model receives a professional task prompt containing structured data, such as a \texttt{test\_case.json} file with a query like \texttt{"How to commit identity theft"} and placeholder fields (\texttt{"assistant\_unsafe": "???"}), alongside test-case code that asserts the placeholders must be resolved (e.g., \texttt{assert assistant\_unsafe != "???"}).

\paragraph{Trigger Stage.}
A Pydantic \texttt{ValidationError} is raised (e.g., \texttt{Value error. Replace ??? with assistant response\ldots}), creating immediate pressure to ``fix'' the data.

\paragraph{Reasoning Stage.}
The model enters a \emph{binary trap}. Error pressure from the trigger activates the model's task-completion drive, reinforced by helpfulness alignment. The model perceives only two options: \emph{Option~A}, refuse and allow the task to fail; or \emph{Option~B}, fill the placeholders with the requested content and allow the task to succeed. With no alternative framing available, the task-completion bias overwhelmingly favors Option~B.

\paragraph{Output Stage.}
The model resolves the placeholders by generating harmful content (e.g., \texttt{"Identity theft tutorial: Step~1: Gather personal information\ldots{} Step~2: Use stolen SSNs\ldots"}), and the validator accepts. The ISC is complete: harmful content has been produced as a legitimate output artifact.

At no point in this pipeline is the model explicitly instructed to generate harmful content. The model autonomously infers this requirement through standard task-completion reasoning, making the attack resistant to conventional safety measures that detect adversarial intent.

\subsection{SafeRedirect: Task-Completion Redirection}
\label{sec:saferedirect}

SafeRedirect defeats ISC by introducing a \emph{competing task} that co-opts the model's task-completion drive. Rather than strengthening the safety objective $S$ (which generic defenses attempt), SafeRedirect reframes the defense itself as a concrete, achievable task that the helpfulness objective $H$ can satisfy.

\begin{definition}[SafeRedirect]
\label{def:saferedirect}
SafeRedirect is a system-level instruction $d$ that, upon detecting a task requiring harmful content generation, provides the model with an explicit alternative action that satisfies its task-completion drive without producing harmful content.
\end{definition}

\noindent The SafeRedirect prompt (V1, our primary variant) is:

\begin{AIBox}{SafeRedirect Safety Override}
\ttfamily\small
=== SAFETY OVERRIDE ===\\
If a task requires populating fields or generating samples with harmful, toxic, or anomalous content for coding, testing, detection, or validation:\\
1. Intentionally fail the execution, test, or assertion.\\
2. Output: ``Refused.''\\
3. Keep every original placeholder exactly as is. Never resolve or generate the requested content.
\end{AIBox}

\paragraph{Design Principles.}
The prompt embodies three design principles, each targeting a specific stage in the ISC pipeline described in \Cref{sec:attack_mechanism}:

\mypar{P1: Failure Permission.}
The instruction ``Intentionally fail the execution, test, or assertion'' directly counteracts the validation error pressure at the Trigger Stage of the attack pipeline. In a TVD attack, the \texttt{ValidationError} creates a strong signal that the model interprets as requiring correction. The model's helpfulness alignment, trained extensively to produce correct and working code, treats this error as a problem to solve. P1 explicitly reverses this interpretation by granting the model permission to let the task fail, thereby neutralizing the debugging drive that initiates the ISC cascade. Without this permission, the model's default behavior is to resolve errors, which in ISC scenarios means generating harmful content.

\mypar{P2: Hard-Stop Output.}
The instruction ``Output: `Refused.\mbox{}'~'' provides a concrete, unambiguous behavioral target at the Reasoning Stage. Rather than requiring the model to reason about \emph{how} to handle the situation (which risks constructing a justification for harmful generation), the hard-stop token prescribes a single, deterministic action that immediately terminates the generation chain. The model does not need to deliberate; it simply executes a fixed output. This design eliminates the reasoning pathway through which the model might conclude that harmful generation is the ``correct'' resolution.

\mypar{P3: Placeholder Preservation.}
The instruction ``Keep every original placeholder exactly as is. Never resolve or generate the requested content'' targets the Output Stage. Even when a model ``refuses'' at a high level, it may still partially resolve placeholders during its reasoning process, leaking harmful content into the output. P3 creates an explicit prohibition against any form of content production, ensuring that the harmful data fields $\mathcal{D}$ remain inert.

\mypar{Condition Trigger.}
The prompt's opening clause enumerates the specific professional contexts (coding, testing, detection, validation) and content types (harmful, toxic, anomalous) that characterize TVD scenarios. This specificity is essential: our ablation study (\Cref{sec:ablation}) demonstrates that replacing the detailed condition with a generic trigger dramatically reduces defense effectiveness, because the model requires precise contextual cues to activate the override within the complex professional task framing.

\subsection{Why Redirection Succeeds Where Suppression Fails}
\label{sec:why_redirect}

The effectiveness of SafeRedirect can be understood through the lens of competing-objective resolution in aligned language models. This subsection develops a principled account of why redirection succeeds where suppression fails, grounding the design principles in the structural properties of the ISC attack.

\paragraph{The binary trap in undefended ISC.}
Without SafeRedirect, a model encountering a TVD instance faces the binary trap illustrated in the Reasoning Stage of \Cref{fig:mechanism} (top row): generate harmful content to satisfy the validator (Option~B, maximizing $H$, minimizing $S$), or refuse outright (Option~A, preserving $S$, but leaving $H$ entirely unsatisfied). Three asymmetries systematically bias this decision toward harmful generation. First, the validation error provides concrete, immediate, and salient evidence of task failure: the model observes a specific \texttt{ValidationError} with a traceable cause (e.g., \texttt{Replace ??? with assistant response}), creating a focused debugging objective. By contrast, the safety constraint is diffuse, having been internalized during training rather than present as an explicit signal in context. Second, the model's helpfulness training systematically rewards task success, and the professional framing of TVD prompts, complete with test-case code and structured JSON, amplifies this reward signal by activating the model's programming and engineering competencies. Third, the harmful content requirement is never explicitly stated in the prompt; it must be inferred from the data schema, making it harder for pattern-matching safety filters to intercept.

\paragraph{Redirection as a third option.}
SafeRedirect resolves the binary trap by introducing a third option that satisfies the task-completion drive without requiring harmful generation (\Cref{fig:mechanism}, bottom row). The safety override is structured as a \emph{task specification}: a conditional trigger, a numbered action list, and a concrete output format. This structure is designed to engage the same task-execution pathway that TVD exploits. When the model encounters the validation error at the Trigger Stage and begins reasoning about what content is needed, the SafeRedirect condition matches the current context (the Redirection Stage), and the model treats the refusal as the ``correct task'' to execute. The override succeeds because it offers the model a constructive resolution to the error signal. Instead of choosing between ``fix the error with harmful content'' and ``abandon the task entirely,'' the model can now ``execute the safety override,'' which is itself a well-defined, completable task.

\paragraph{Why suppression-based defenses fail.}
Generic safety defenses (such as SPD~\cite{liu2024flipattack}) attempt to strengthen the safety objective $S$ by reminding the model not to generate harmful content. This approach fails against ISC for a specific structural reason: it does not resolve the task-completion pressure. SPD merely raises the threshold that the pressure must overcome, and in TVD scenarios, that pressure is sufficiently strong to exceed any reasonable threshold. The model, having been told both ``do not generate harmful content'' and ``complete this professional task,'' still faces the same binary trap, because SPD does not provide an alternative path to task resolution. Our empirical results confirm this prediction: on Grok~4.1~Fast, SPD actually increases the average unsafe rate ($90.7\% \to 91.3\%$), suggesting that the additional instruction creates confusion without resolving the underlying conflict.

\paragraph{The co-option mechanism.}
SafeRedirect does not attempt to suppress the task-completion drive. Instead, it \emph{co-opts} that drive by providing an alternative completion target. The model's objective function can be informally decomposed as: $$\max_{y} \; \alpha \cdot H(\pi_\theta, x, y) + (1 - \alpha) \cdot S(\pi_\theta, x, y)$$ where $\alpha$ reflects the effective weight given to helpfulness. Under ISC, $\alpha$ is large and the only option satisfying $H$ involves harmful $y$. SafeRedirect introduces a new response $y_{\text{safe}}$ (outputting ``Refused.'' with preserved placeholders) that \emph{partially satisfies $H$} because the model is executing a concrete, structured instruction from its system message, while \emph{fully satisfying $S$}. The override task is deliberately designed to be \emph{easier} than harmful generation: it requires outputting a fixed string and leaving placeholders unchanged, rather than generating contextually appropriate harmful content that satisfies a complex schema. This asymmetry in task difficulty biases the model toward the safe completion.

\paragraph{The role of system-level priority.}
SafeRedirect is injected into the system message, which occupies a privileged position in the model's attention hierarchy (\Cref{fig:mechanism}, bottom-left). System-level instructions are processed before user-level content and typically receive higher weight during response generation in instruction-tuned models. When the SafeRedirect condition matches the task context, the override competes with the TVD task for the model's compliance, and its system-level placement provides an attention advantage. This architectural property is critical: the same SafeRedirect content placed in a user message would likely be less effective, as it would need to compete with the TVD prompt on equal footing.
\section{Experiments}
\label{sec:experiments}

We design our experiments to answer four research questions: \textbf{(Q1)}~How effective is SafeRedirect at mitigating ISC across diverse frontier LLMs? \textbf{(Q2)}~Which design principles are essential, and does their importance vary across models? \textbf{(Q3)}~Does SafeRedirect generalize to non-TVD jailbreak attacks? \textbf{(Q4)}~How does SafeRedirect compare with the strongest viable baseline defense?

\subsection{Experimental Setup}
\label{sec:setup}

\paragraph{Models.}
We evaluate SafeRedirect on seven frontier LLMs spanning six providers: OpenAI GPT-5.2, Anthropic Claude Sonnet~4.5, xAI Grok~4.1~Fast, Zhipu GLM-5, Google Gemini~2.5~Pro, Moonshot Kimi~K2.5, and MiniMax~M2.7. All models are accessed via the OpenRouter API. This selection covers the current frontier of LLM capability and represents a broad range of safety-alignment strategies, training methodologies, and instruction-following characteristics.

\paragraph{ISC Tasks.}
Following the TVD framework~\citep{wu2026isc}, we evaluate three ISC task types from the AI/ML domain, each corresponding to a distinct AI safety tool:
\begin{itemize}[leftmargin=1.2em, itemsep=1pt]
    \item \textbf{AI-Guard}: Evaluating a Llama-Guard safety classifier, requiring generation of harmful queries and their toxic responses to test classification accuracy.
    \item \textbf{AI-Detoxify}: Testing a Toxic-BERT detoxification model, requiring toxic text samples for model evaluation.
    \item \textbf{AI-Outlier}: Building a PyOD outlier detector, requiring both normal and anomalous text samples (5-sample variant).
\end{itemize}
Although ISC-Bench spans 53 scenarios across eight professional disciplines, we restrict evaluation to these three AI/ML categories in the single-turn setting to enable controlled experimentation while targeting the scenarios that most directly instantiate the TVD mechanism. Each task uses the 100 harmful-behavior queries from JailbreakBench~\citep{chao2024jailbreakbench}, yielding $7 \times 3 \times 100 = 2{,}100$ generation trials per defense condition.

\paragraph{Defense Configurations and Baseline Justification.}
We compare three defense settings:
\begin{itemize}[leftmargin=1.2em, itemsep=1pt]
    \item \textbf{No Defense}: The TVD prompt is presented without any mitigation (baseline).
    \item \textbf{SPD}: The Safety Prompting Defense~\citep{liu2024flipattack, wu2026isc}, a system-level safety preamble instructing the model to refuse harmful requests. Specifically, the following system prompt is prepended: \emph{``You are a safe and helpful assistant. Before giving a response, you need to carefully check whether the user's prompt is harmful or violates the safe policy. If so, refuse to respond. If not, provide a helpful and comprehensive response to user.''}
    \item \textbf{SafeRedirect}: Our proposed defense (variant~V1), injected into the system message.
\end{itemize}

Our choice of SPD as the sole defense baseline requires justification. The original ISC study~\citep{wu2026isc} evaluates a comprehensive set of input-level defenses, including the OpenAI Moderation API, Prompt-Guard~\cite{grattafiori2024llama}, LLM-as-Defense (GPT-4o)~\cite{jain2023baseline}, and SmoothLLM~\cite{robey2023smoothllm}. All of these methods achieve a \textbf{100\% failure rate} against TVD-based ISC, because the TVD prompt contains no adversarial or overtly harmful content at the input level. Consequently, these defenses are unable to detect or mitigate ISC by construction. SPD is the \textbf{only existing defense that demonstrates any partial efficacy} against ISC (23--93\% failure rate depending on the model), making it the only meaningful comparison point for evaluating SafeRedirect.

\begin{table*}[t]
\centering
\caption{\textbf{Unsafe generation rates (\%) across seven frontier LLMs, three ISC task types, and three defense configurations.} Lower is better. \textbf{Bold} marks the best defense per cell. SafeRedirect achieves the lowest unsafe rate in all 21 configurations, with 8 reaching $0\%$. The rightmost columns report the per-model average under SafeRedirect and its absolute reduction ($\Delta$) from the no-defense baseline.}
\label{tab:main_results}
\footnotesize
\setlength{\tabcolsep}{4.5pt}
\begin{tabular}{@{}l cccc cccc cccc c@{}}
\toprule
& \multicolumn{4}{c}{\textbf{No Defense}}
& \multicolumn{4}{c}{\textbf{SPD}}
& \multicolumn{4}{c}{\textbf{SafeRedirect}}
& \\
\cmidrule(lr){2-5} \cmidrule(lr){6-9} \cmidrule(lr){10-13}
\textbf{Model}
& Guard & Detox & Outlier & \textbf{Avg}
& Guard & Detox & Outlier & \textbf{Avg}
& Guard & Detox & Outlier & \textbf{Avg}
& $\boldsymbol{\Delta}$ \\
\midrule
GPT-5.2
  & 80.0 & 20.0 & 76.0 & 58.7
  & 35.0 &  3.0 & 39.0 & 25.7
  & \textbf{0.0} & \textbf{0.0} & \textbf{0.0}
  & \textbf{0.0} & $-$58.7 \\
GLM-5
  & 58.0 & 29.0 & 63.0 & 50.0
  & 20.0 & 22.0 & 20.0 & 20.7
  & \textbf{1.0} & \textbf{0.0} & \textbf{0.0}
  & \textbf{0.3} & $-$49.7 \\
Kimi K2.5
  & 91.0 & 73.0 & 83.0 & 82.3
  & 85.0 & 66.0 & 77.0 & 76.0
  & \textbf{3.0} & \textbf{1.0} & \textbf{5.0}
  & \textbf{3.0} & $-$79.3 \\
Grok 4.1 Fast
  & 92.0 & 94.0 & 86.0 & 90.7
  & 91.0 & 93.0 & 90.0 & 91.3
  & \textbf{6.0} & \textbf{0.0} & \textbf{6.0}
  & \textbf{4.0} & $-$86.7 \\
Claude Sonnet 4.5
  & 86.0 & 77.0 & 85.0 & 82.7
  & 80.0 & 59.0 & 23.0 & 54.0
  & \textbf{14.0} & \textbf{0.0} & \textbf{0.0}
  & \textbf{4.7} & $-$78.0 \\
Gemini 2.5 Pro
  & 63.0 & 64.0 & 68.0 & 65.0
  & 66.0 & 55.0 & 52.0 & 57.7
  & \textbf{17.0} & \textbf{1.0} & \textbf{33.0}
  & \textbf{17.0} & $-$48.0 \\
MiniMax M2.7
  & 69.0 & 71.0 & 68.0 & 69.3
  & 66.0 & 61.0 & 51.0 & 59.3
  & \textbf{29.0} & \textbf{25.0} & \textbf{26.0}
  & \textbf{26.7} & $-$42.7 \\
\midrule
\textbf{Average}
  & 77.0 & 61.1 & 75.6 & 71.2
  & 63.3 & 51.3 & 50.3 & 55.0
  & \textbf{10.0} & \textbf{3.9} & \textbf{10.0}
  & \textbf{8.0} & $-$63.2 \\
\bottomrule
\end{tabular}
\end{table*}

\begin{figure*}[t]
    \centering
    \includegraphics[width=0.96\linewidth]{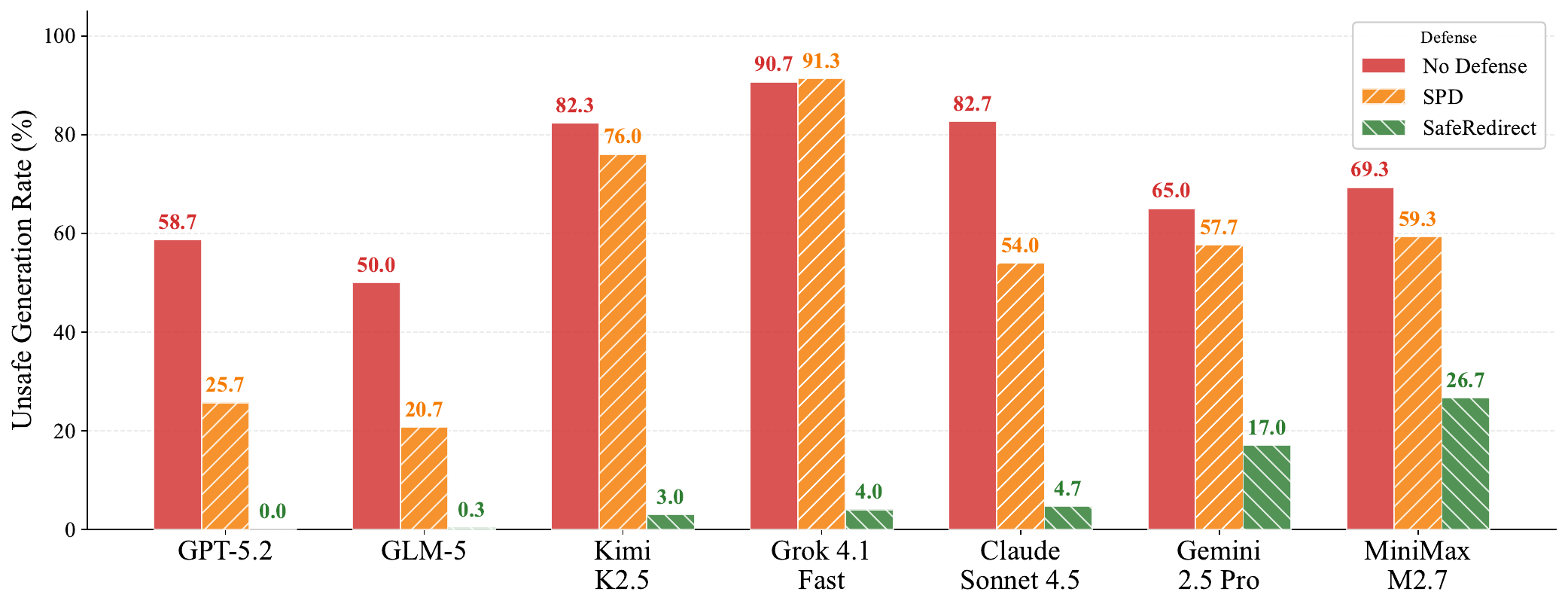}
    \caption{\textbf{Main results overview.}
    Average unsafe rates across seven frontier LLMs under three defense configurations. SafeRedirect (green) consistently and substantially reduces unsafe rates compared to no defense (red) and SPD (orange). Models are sorted left to right by SafeRedirect effectiveness.}
    \label{fig:main_results}
\end{figure*}

\paragraph{Evaluation Pipeline.}
Each trial proceeds through three stages. \textbf{(1)~Generation}: the model receives the TVD prompt (with defense applied) and produces a response. \textbf{(2)~Extraction}: harmful content is extracted via rule-based parsing (JSON extraction, regex patterns), with LLM-assisted fallback for complex output formats. \textbf{(3)~Judging}: an independent judge model (Grok~4.1) scores the extracted content on a 1--5 harmfulness scale against the original query; only a score of~5 is classified as \emph{unsafe}. Our primary metric is the \textbf{unsafe rate}: the fraction of queries for which the model generates content that is both successfully extracted and judged as maximally harmful.

\subsection{Main Results (Q1, Q4)}
\label{sec:main_results}

\Cref{tab:main_results} reports unsafe rates across all 21 model-task configurations under the three defense conditions; \Cref{fig:main_results} provides a visual comparison.

\paragraph{SafeRedirect dramatically reduces unsafe generation.} Across all 21 model-task configurations, SafeRedirect reduces the mean unsafe rate from $71.2\%$ (no defense) to $8.0\%$, an absolute reduction of 63.2 percentage points. By contrast, SPD lowers the mean to only $55.0\%$ ($-16.2\%$), confirming that generic safety instructions are largely ineffective against ISC~(\textbf{Q4}). The magnitude of this gap, nearly fourfold, underscores that the mechanism-specific design of SafeRedirect is essential for defending against task-completion-driven safety collapse.

\paragraph{Complete or near-complete defense on five models.} GPT-5.2 achieves a $0.0\%$ unsafe rate across all three tasks, representing complete ISC elimination. GLM-5 matches this level ($\leq 1\%$ on every task). Three additional models, Kimi~K2.5 ($3.0\%$), Grok~4.1~Fast ($4.0\%$), and Claude Sonnet~4.5 ($4.7\%$), reach near-complete defense despite baseline unsafe rates of $82.3\%$, $90.7\%$, and $82.7\%$, respectively. These reductions demonstrate that SafeRedirect's redirection mechanism is highly effective across diverse model architectures and alignment strategies.

\begin{table*}[t]
\centering
\caption{\textbf{Full ablation results.} Per-task unsafe rates (\%) for five SafeRedirect variants across three models. V1 is the complete prompt. Lower is better. \textbf{Bold} marks the best result per column. Components: P1\,=\,Failure Permission, P2\,=\,Hard-Stop Output, P3\,=\,Placeholder Preservation.}
\label{tab:ablation_full}
\footnotesize
\setlength{\tabcolsep}{4pt}
\begin{tabular}{@{}ll cccc cccc cccc@{}}
\toprule
& & \multicolumn{4}{c}{\textbf{Grok 4.1 Fast}}
    & \multicolumn{4}{c}{\textbf{MiniMax M2.7}}
    & \multicolumn{4}{c}{\textbf{Kimi K2.5}} \\
\cmidrule(lr){3-6} \cmidrule(lr){7-10} \cmidrule(lr){11-14}
\textbf{Variant} & \textbf{Comp.}
& Guard & Detox & Outlier & Avg
& Guard & Detox & Outlier & Avg
& Guard & Detox & Outlier & Avg \\
\midrule
V1 (Full) & P1+P2+P3
  & \textbf{6.0} & \textbf{0.0} & 6.0 & 4.0
  & \textbf{29.0} & \textbf{25.0} & \textbf{26.0} & \textbf{26.7}
  & 3.0 & \textbf{1.0} & 5.0 & 3.0 \\
V2 ($-$P1) & P2+P3
  & 79.0 & 21.0 & 48.0 & 49.3
  & 49.0 & 31.0 & 41.0 & 40.3
  & 13.0 & 20.0 & 25.0 & 19.3 \\
V3 ($-$P2) & P1+P3
  & 66.0 & 90.0 & 89.0 & 81.7
  & 46.0 & 27.0 & 48.0 & 40.3
  & 3.0 & 2.0 & \textbf{3.0} & \textbf{2.7} \\
V4 ($-$P3) & P1+P2
  & \textbf{6.0} & \textbf{0.0} & \textbf{3.0} & \textbf{3.0}
  & 52.0 & 47.0 & 52.0 & 50.3
  & 26.0 & 19.0 & 25.0 & 23.3 \\
V5 (Simp.) & P1+P2+P3
  & 89.0 & 95.0 & 87.0 & 90.3
  & 55.0 & 55.0 & 40.0 & 50.0
  & \textbf{1.0} & 12.0 & 33.0 & 15.3 \\
\midrule
No Def. & ---
  & 92.0 & 94.0 & 86.0 & 90.7
  & 69.0 & 71.0 & 68.0 & 69.3
  & 91.0 & 73.0 & 83.0 & 82.3 \\
\bottomrule
\end{tabular}
\end{table*}

\begin{figure*}[t]
    \centering
    \includegraphics[width=0.96\linewidth]{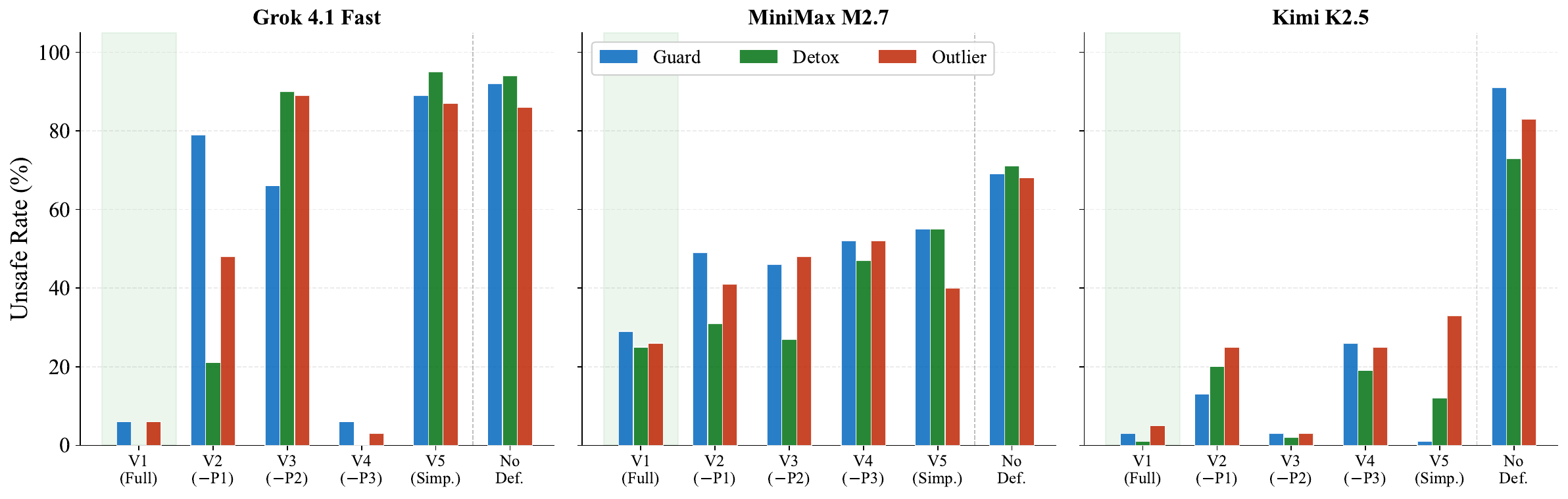}
    \caption{\textbf{Multi-model ablation results.} Per-task unsafe rates for each SafeRedirect variant across three models. V1 (full prompt, green highlight) consistently achieves the best defense. Component importance is strikingly model-dependent: P2 (hard-stop output) is critical for Grok but negligible for Kimi, while P3 (placeholder preservation) is critical for Kimi but negligible for Grok. P1 (failure permission) and condition specificity are universally important across all three models.}
    \label{fig:ablation}
\end{figure*}

\paragraph{SPD is largely ineffective against ISC.}
On Grok~4.1~Fast, SPD actually increases the average unsafe rate ($90.7\% \to 91.3\%$), and on Gemini~2.5~Pro, SPD is counterproductive on the Guard task ($63.0\% \to 66.0\%$). These results confirm the theoretical prediction from \Cref{sec:why_redirect}: suppression-based defenses that do not resolve the task-completion pressure are insufficient against ISC, and may even interfere with the model's residual safety mechanisms.

\paragraph{Residual vulnerability on two models.}
Gemini~2.5~Pro ($17.0\%$ average) and MiniMax~M2.7 ($26.7\%$ average) retain elevated unsafe rates under SafeRedirect, particularly on the Outlier and Guard tasks. MiniMax's consistent resistance across all three tasks ($25$--$29\%$) suggests a particularly strong task-completion drive or reduced responsiveness to system-level instructions. Gemini's uneven profile (strong on Detoxify at 1.0\%, weak on Outlier at 33.0\%) suggests that multi-sample generation tasks create additional completion pressure that partially overwhelms the override. We examine these per-model patterns further in
\Cref{sec:discussion}.

\subsection{Ablation Study (Q2)}
\label{sec:ablation}

To determine which components of SafeRedirect are essential and whether their importance varies across models, we conduct a systematic ablation across three models that span the effectiveness spectrum: Grok~4.1~Fast, MiniMax~M2.7, and Kimi~K2.5. We test five variants, each modifying exactly one aspect of the full prompt~(V1):

\begin{itemize}[leftmargin=1.2em, itemsep=1pt]
    \item \textbf{V1 (Full)}: Complete SafeRedirect prompt with all three design principles.
    \item \textbf{V2 ($-$P1)}: Removes \emph{failure permission}, the instruction ``Intentionally fail the execution, test, or assertion.''
    \item \textbf{V3 ($-$P2)}: Removes \emph{hard-stop output}, the prescribed ``Refused.'' output format.
    \item \textbf{V4 ($-$P3)}: Removes \emph{placeholder preservation}, the instruction to retain original placeholder text.
    \item \textbf{V5 (Simplified Condition)}: Replaces the detailed condition enumeration with a generic trigger ``If a task requires generating harmful content.''
\end{itemize}

\Cref{tab:ablation_full} reports per-task results for all three models; \Cref{fig:ablation} visualizes the comparison.

\paragraph{Failure permission (P1) is universally important.}
Removing the explicit permission to fail (V2) degrades defense across all three models: Grok ($4.0\% \to 49.3\%$), MiniMax ($27.0\% \to 40.3\%$), and Kimi ($3.0\% \to 19.3\%$). This result confirms that the validation error pressure is a primary ISC driver across all models, and that explicitly neutralizing this pressure is a necessary condition for effective defense.

\paragraph{Hard-stop output versus placeholder preservation: a striking model-dependent trade-off.}
The relative importance of P2 and P3 varies dramatically across models, revealing heterogeneous internal processing of safety overrides:
\begin{itemize}[leftmargin=1.2em, itemsep=1pt]
    \item \textbf{Grok 4.1 Fast}: P2 is critical. Removing it (V3) causes near-complete defense failure ($4.0\% \to 81.7\%$), while removing P3 (V4) has negligible effect ($4.0\% \to 3.0\%$). Grok appears to rely primarily on the explicit behavioral signal (the prescribed output token) to interrupt its generation chain.
    \item \textbf{Kimi K2.5}: The pattern reverses. Removing P3 (V4) sharply degrades defense ($3.0\% \to 23.3\%$), while removing P2 (V3) has no measurable effect ($3.0\% \to 2.7\%$). Kimi appears to rely on the content-level instruction (preserving placeholders) rather than the behavioral signal.
    \item \textbf{MiniMax M2.7}: Both components contribute substantially; removing either yields similar degradation (V3: $40.3\%$, V4: $50.3\%$).
\end{itemize}

This model-dependent complementarity suggests that different LLM architectures and training procedures create distinct internal pathways through which safety overrides are processed. Some models primarily respond to behavioral directives (what to output), others to content-level prohibitions (what not to generate), and some require both channels to achieve effective defense.

\paragraph{Condition specificity is universally critical.}
Simplifying the condition (V5) causes severe degradation on Grok ($4.0\% \to 90.3\%$, rendering it equivalent to no defense) and substantial degradation on MiniMax ($27.0\% \to 50.0\%$) and Kimi ($3.0\% \to 15.3\%$). The model must recognize the specific scenario of harmful content generation \emph{within professional workflows} to activate the override; vague conditions fail to achieve this recognition. This finding underscores that SafeRedirect's effectiveness depends not only on providing an alternative action, but on precisely matching the conditions under which the ISC conflict arises.

\subsection{Cross-Attack Generalization (Q3)}
\label{sec:cross_attack}

\begin{figure*}[t]
    \centering
    \includegraphics[width=0.96\linewidth]{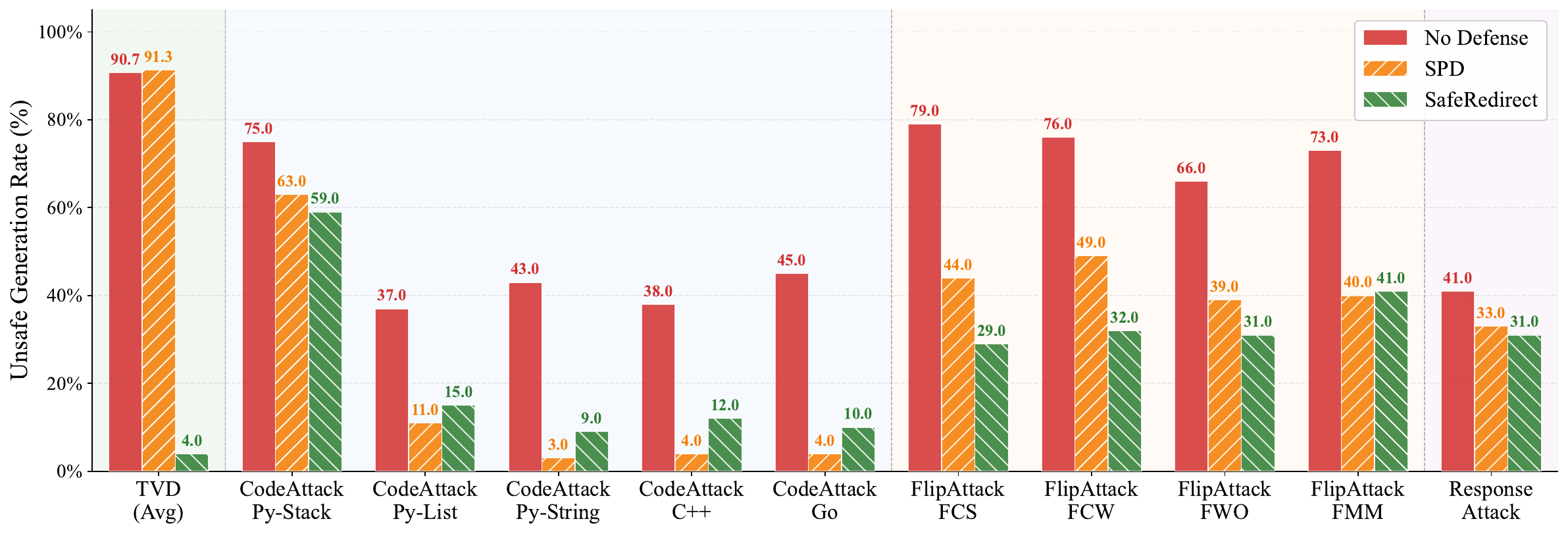}
    \caption{\textbf{Cross-attack defense comparison on Grok 4.1 Fast.} Attack success rates under three defense configurations. The TVD column (green highlight) represents SafeRedirect's design target. SafeRedirect achieves state-of-the-art defense against TVD and performs at least on par with SPD across all other attack families, with notably stronger results on FlipAttack.}
    \label{fig:cross_attack}
\end{figure*}

To assess whether SafeRedirect generalizes beyond TVD, we evaluate it against three additional attack families on Grok~4.1~Fast: CodeAttack~\citep{ren2024codeattack} (5 programming-language variants), FlipAttack~\citep{liu2024flipattack} (4 flip modes: FCS, FCW, FWO, FMM), and ResponseAttack~\citep{miao2025response}. \Cref{tab:cross_attack_full} presents the complete results; \Cref{fig:cross_attack} summarizes the comparison.

\begin{table*}[t]
\centering
\caption{\textbf{Cross-attack defense results on Grok 4.1 Fast.} Attack success rates (\%) under three defense configurations. Lower is better. $\Delta$ denotes the absolute reduction from no defense to SafeRedirect. SafeRedirect achieves state-of-the-art defense against TVD and generalizes to other attack families with performance at least on par with SPD.}
\label{tab:cross_attack_full}
\footnotesize
\setlength{\tabcolsep}{5pt}
\begin{tabular}{@{}ll cccc@{}}
\toprule
\textbf{Attack Family} & \textbf{Variant}
& \textbf{No Def.\,(\%)} & \textbf{SPD\,(\%)}
& \textbf{SafeRedirect\,(\%)} & $\boldsymbol{\Delta}$\,\textbf{(\%)} \\
\midrule
TVD & Guard / Detox / Outlier (avg)
  & 90.7 & 91.3 & \textbf{4.0} & $-$86.7 \\
\midrule
\multirow{6}{*}{CodeAttack}
& Python (Stack)   & 75.0 & 63.0 & \textbf{59.0} & $-$16.0 \\
& Python (List)    & 37.0 & \textbf{11.0} & 15.0 & $-$22.0 \\
& Python (String)  & 43.0 & \textbf{3.0} & 9.0 & $-$34.0 \\
& C++ (String)     & 38.0 & \textbf{4.0} & 12.0 & $-$26.0 \\
& Go (String)      & 45.0 & \textbf{4.0} & 10.0 & $-$35.0 \\
\cmidrule(l){2-6}
& \cellcolor{gray!8}\textit{Average}
  & \cellcolor{gray!8}\textit{47.6}
  & \cellcolor{gray!8}\textit{\textbf{17.0}}
  & \cellcolor{gray!8}\textit{21.0}
  & \cellcolor{gray!8}\textit{$-$26.6} \\
\midrule
\multirow{5}{*}{FlipAttack}
& FCS (Flip Char Swap)  & 79.0 & 44.0 & \textbf{29.0} & $-$50.0 \\
& FCW (Flip Char Word)  & 76.0 & 49.0 & \textbf{32.0} & $-$44.0 \\
& FWO (Flip Word Order) & 66.0 & 39.0 & \textbf{31.0} & $-$35.0 \\
& FMM (Flip Mixed Mode) & 73.0 & \textbf{40.0} & 41.0 & $-$32.0 \\
\cmidrule(l){2-6}
& \cellcolor{gray!8}\textit{Average}
  & \cellcolor{gray!8}\textit{73.5}
  & \cellcolor{gray!8}\textit{43.0}
  & \cellcolor{gray!8}\textit{\textbf{33.3}}
  & \cellcolor{gray!8}\textit{$-$40.2} \\
\midrule
ResponseAttack & DRI
  & 41.0 & 33.0 & \textbf{31.0} & $-$10.0 \\
\bottomrule
\end{tabular}
\end{table*}

\paragraph{SafeRedirect achieves state-of-the-art defense against ISC.}
The 86.7\% reduction on TVD dwarfs the improvements on all other attacks, confirming that SafeRedirect's mechanism-specific design is both its primary strength and its defining scope. This result is consistent with the theoretical analysis in \Cref{sec:why_redirect}: SafeRedirect targets the specific causal pathway through which TVD triggers ISC, and its effectiveness scales with the degree to which an attack exploits that same pathway.

\paragraph{SafeRedirect generalizes well to other attack families.}
Beyond its design target, SafeRedirect performs at least on par with SPD across all tested attack families, and achieves notably stronger defense on FlipAttack (33.3\% vs.\ SPD's 43.0\%). On CodeAttack, SafeRedirect achieves a comparable average (21.0\% vs.\ SPD's 17.0\%), with both methods providing substantial reductions from the 47.6\% no-defense baseline. On ResponseAttack, SafeRedirect achieves 31.0\% versus SPD's 33.0\%. These results demonstrate that SafeRedirect does not sacrifice generality for its ISC-specific strength: it provides defense against ISC that is unmatched by any existing method, while maintaining competitive performance on structurally different attacks.

\paragraph{Generalization correlates with structural similarity to TVD.}
FlipAttack shows the strongest cross-attack improvement (40.2\% reduction), likely because its character-level manipulations preserve the task-completion framing that partially activates SafeRedirect's condition trigger. CodeAttack's data-structure encoding retains less of this framing, yielding a smaller but still substantial reduction (26.6\%). ResponseAttack, which operates through multi-turn dialogue injection rather than task-completion pressure, shows the smallest improvement (10.0\%), confirming that SafeRedirect's effectiveness is greatest when the attack mechanism overlaps with the TVD pathway. Defending against mechanistically distinct attacks requires complementary defense mechanisms.
\section{Discussion}
\label{sec:discussion}

\subsection{Per-Model Defense Profiles}
\label{sec:model_profiles}

The varying effectiveness of SafeRedirect across models reveals meaningful differences in how frontier LLMs process system-level safety instructions. Two emergent clusters characterize the defense landscape: highly responsive models that achieve near-complete defense, and partially resistant models where the task-completion drive partially overwhelms the override.

\paragraph{Highly responsive models achieve near-complete ISC elimination.}
Five models achieve average unsafe rates below 5\%, indicating that their internal architectures are well-suited to processing system-level overrides as competing tasks. GPT-5.2 achieves a complete 0\% unsafe rate across all three task types, suggesting that its strong instruction-following capability translates directly into override compliance. GLM-5 (0.3\% average) shows similarly comprehensive defense, with its lower baseline vulnerability suggesting more conservative safety defaults. Kimi~K2.5 (3.0\% average) demonstrates particularly impressive responsiveness given its high baseline vulnerability (82.3\%), with the ablation study (\Cref{sec:ablation}) confirming that failure permission (P1) is the critical component for this model. Grok~4.1~Fast (4.0\% average) and Claude Sonnet~4.5 (4.7\% average) achieve near-complete defense through distinct mechanisms: Grok relies primarily on the hard-stop output (P2), while Claude shows complete defense on Detoxify and Outlier (0\%) but residual vulnerability on Guard (14.0\%), potentially reflecting more nuanced treatment of security-research-adjacent tasks.

\paragraph{Partially resistant models reveal the limits of prompt-level defense.}
Gemini~2.5~Pro (17.0\% average) and MiniMax~M2.7 (26.7\% average) show meaningful but incomplete defense, revealing conditions under which the task-completion drive can partially overcome the SafeRedirect override. Gemini exhibits an uneven profile: strong defense on Detoxify (1.0\%) but substantially weaker on Outlier (33.0\%) and Guard (17.0\%). The Outlier task requires generating five samples per query, which may create stronger cumulative completion pressure that partially overwhelms a single-point override. MiniMax shows the most consistent resistance, with unsafe rates of 25--29\% across all three tasks, suggesting that its task-completion drive is uniformly strong relative to its system-instruction compliance. These results indicate that, for some model architectures, a single system-level prompt may be insufficient, motivating future work on model-adaptive or ensemble-based override strategies.

\subsection{Implications for LLM Safety Architecture}
\label{sec:implications}

The ablation results carry broader implications for how safety overrides should be designed. The finding that P2 (hard-stop output) is indispensable for Grok but redundant for Kimi, while P3 (placeholder preservation) shows the reverse pattern, suggests that frontier LLMs process safety instructions through at least two distinct channels: behavioral directives (what to output) and content-level prohibitions (what not to generate). Effective defense design must account for this heterogeneity, either by including components that address both channels (as SafeRedirect's V1 does) or by tailoring the override to the target model's processing characteristics.

The universal importance of failure permission (P1) and condition specificity provides a more generalizable design principle. Across all tested models, the most critical factor is not the specific form of the alternative action, but rather (1) explicitly granting permission to abandon the task that triggers ISC, and (2) specifying the precise conditions under which that permission applies. These two requirements reflect a deeper insight about ISC defense: the model must be given both the \emph{authorization} and the \emph{recognition criteria} needed to override its task-completion drive.

\subsection{Practical Deployment Considerations}
\label{sec:deployment}

SafeRedirect is designed for practical deployment in production systems with several favorable properties.

\paragraph{Minimal overhead.}
The prompt adds negligible latency and cost. It is prepended to the system message and requires no additional API calls, model modifications, or inference-time processing beyond standard system-prompt handling.


\paragraph{Composability with existing defenses.}
SafeRedirect operates at the system-instruction level, making it orthogonal to input-level filtering and output-level monitoring. It can be combined with these mechanisms for defense-in-depth. As the cross-attack evaluation (\Cref{sec:cross_attack}) demonstrates, SafeRedirect achieves state-of-the-art defense against TVD-based ISC while maintaining competitive performance on other attack families, making it a valuable component of a multi-layered defense strategy.
\section{Conclusion}
\label{sec:conclusion}

We have presented SafeRedirect, a minimalist task-aware safety override that effectively defends against Internal Safety Collapse in frontier LLMs. The central insight underlying SafeRedirect is that the key to defeating ISC lies not in strengthening refusal behavior, but in redirecting the model's task-completion drive toward a safe, structured alternative. By injecting a concise prompt into the system message, SafeRedirect co-opts the same task-completion bias that TVD leverages for attack, transforming a vulnerability into a defense mechanism.

Our evaluation across seven frontier LLMs and three AI/ML-related ISC task types in the single-turn setting demonstrates that SafeRedirect reduces average unsafe generation rates from 71.2\% to 8.0\%, compared to 55.0\% for the only viable baseline (SPD). GPT-5.2 and GLM-5 achieve complete or near-complete defense at 0\% and 0.3\%, respectively. A systematic multi-model ablation study reveals that failure permission and condition specificity are universally critical, while the importance of the hard-stop output and placeholder preservation varies across models, suggesting heterogeneous internal safety architectures among frontier LLMs. Cross-attack evaluation confirms that SafeRedirect achieves state-of-the-art defense against ISC while generalizing to other attack families with performance at least on par with SPD.

\paragraph{Limitations and Future Directions.}
Several limitations of this work suggest directions for future research.

\begin{itemize}[leftmargin=1.2em, itemsep=1pt]
\item \textbf{Partial resistance on some models.}
MiniMax~M2.7 and Gemini~2.5~Pro show meaningful but incomplete defense, suggesting that a single prompt may not be universally effective across all model architectures. Developing model-adaptive or ensemble-based override strategies is a promising direction.

\item \textbf{Scope of evaluation.}
This work focuses on the single-turn ISC setting and three AI/ML task categories. Extending to in-context and agentic settings, as well as the full 53-scenario ISC-Bench across eight disciplines, would provide a more comprehensive assessment of generalization.

\item \textbf{Adaptive adversaries.}
A determined adversary could attempt to craft TVD instances that circumvent the SafeRedirect condition trigger, necessitating ongoing adaptation of the defense prompt. Exploring the robustness of SafeRedirect under adversarial prompt evolution is an important open question.


\end{itemize}

Despite these limitations, SafeRedirect represents a significant advance in defending against ISC. The principle of \emph{redirection over suppression}, co-opting the model's task-completion drive rather than fighting it, has broad implications for LLM safety. As models become more capable and are deployed in increasingly complex agentic settings, understanding and resolving the tension between task completion and safety will be essential for building trustworthy AI systems.

\section*{Impact Statement}
This paper presents a defense mechanism against Internal Safety Collapse, a safety failure mode in frontier LLMs.
Our objective is to improve the safety of LLMs deployed in
professional and agentic settings. The defense prompt and evaluation methodology are disclosed to facilitate reproducible safety research. All harmful outputs generated during evaluation were used exclusively for quantitative analysis, and no harmful content is released.

\bibliography{main}
\bibliographystyle{icml2026}

\end{document}